\documentclass[preprint,aps,prc,nofootinbib]{revtex4}

\usepackage{epsfig}
\usepackage{graphicx}

\begin{document}
\title{New possibility for further measurements of nucleon form factors at large momentum transfer in time-like region: 
$\overline{p}+{p}\to \ell^+ +\ell^-$, $\ell=e$ or $\mu$.}
\author{Egle Tomasi-Gustafsson}
\email{etomasi@cea.fr}
\affiliation{\it DAPNIA/SPhN, CEA/Saclay, 91191 Gif-sur-Yvette Cedex,
France}
\author{Michail P. Rekalo }
\affiliation{\it National Science Center KFTI, 310108 Kharkov, Ukraine}
\begin{abstract}
We briefly summarize the status with electromagnetic nucleon form factors, and give in this framework, arguments to study the angular dependence of the differential cross section and single-spin polarization phenomena (polarized target or polarized beam) in $\overline{p}+{p}\to \ell^++\ell^-$, in view of the availability of future antiproton beams.
\end{abstract}
\maketitle
Nucleon electromagnetic form factors (FFs) are fundamental quantities for the understanding of the nucleon structure. They can be experimentally measured as well as theoretically calculated and therefore constitute a privileged background for the test of theoretical models. Since their first determination, which valued the Nobel Prize to R. Hofstadter in 1961, new and exciting developments in this field have been recently done, thanks to the advent, of high intensity, highly polarized electron beams, hadron polarimeters and polarized targets, in the space-like region and high energy antiproton beams, at Fermi-Lab.

Recent experimental data on nucleon electromagnetic form factors (FFs) in 
time-like (TL) \cite{An03} and space-like (SL) \cite{Jo00,Ga02} regions of momentum transfer 
square $q^2$, and many new theoretical 
developments show the necessity of a global description of form 
factors in the full region of $q^2$. 

Form factors are analytical functions of $q^2$, 
being real functions in the SL  region (due to the hermiticity of the 
electromagnetic Hamiltonian) and complex functions in the 
TL region. The Phr\`agmen-Lindel\"of 
theorem \cite{Ti39} gives a rigorous prescription for the asymptotic behavior of analytical functions: 
$\lim_{q^2\to -\infty} F^{(SL)}(q^2) =\lim_{q^2\to \infty} 
F^{(TL)}(q^2)$.
This means that, asymptotically, the FFs, have the following constrains: 
1) the time-like phase vanishes  and 2) the real part of the FFs, 
${\cal R}e  F^{(TL)}(q^2)$, coincides with the 
corresponding value $F^{(SL)}(q^2)$.

The existing experimental data about the electromagnetic FFs of charged pion or 
proton in the time-like region do not allow a complete test of   the 
Phr\`agmen-Lindel\"of theorem, especially concerning the vanishing phase, as the cross section depends on the square of the modulus of the form factor. Only the 
study of more complicated reactions such 
as $\pi^-+p\to n+\ell^++\ell^-$ \cite{Re66} or $\overline{p}+p\to \pi^0+\ell^++\ell^-$ 
\cite{Du96} allows, in 
principle: to determine the nucleon FFs in the unphysical region of TL momentum 
transfer, 
for 4$m_e^2 \le q^2\le$ 4$m^2$, where $m_e$ is the leptonic mass, and to determine the relative phase of pion and nucleon form factors.

The cross section for the process $\overline{p}+p\to \ell^++ \ell^-$, for the one-photon mechanism, neglecting the leptonic mass, can be expressed as a function of 
FFs according to the following formula \cite{Zi62}:
\begin{equation}
\displaystyle\frac{d\sigma}{d(cos\theta)}=
\displaystyle\frac{\pi\alpha^2}{8m^2\tau\sqrt{\tau(\tau-1)}}
\left [ \tau |G_M|^2(1+\cos^2\theta)+|G_E|^2\sin^2\theta
\right ],
\label{eq:eqcs}
\end{equation}
where $\theta$ is the angle between the lepton and the antiproton
in the center of 
mass frame, $\tau=q^2/(4m^2)$, $m$ is the nucleon mass, $\alpha=e^2/(4\pi)=1/137$. 

The Rosenbluth 
separation  of the $|G_E|^2$ and $|G_M|^2$ contributions, in TL region, which is equivalent here to the linearity of $d\sigma/d\cos\theta$, with respect to $\cos^2\theta$, has not been realized yet.

In order to determine the form factors, due to the poor statistics in the existing data \cite{An03}, it 
is necessary to integrate the differential cross section over a wide angular 
range. One typically assumes that the $G_E$-contribution plays a minor role in the cross 
section at large $q^2$ and the 
experimental results are usually given 
in terms of $|G_M|$, under the hypothesis that $G_E=0$ or $|G_E|=|G_M|$. The first hypothesis is arbitrary. The second hypothesis is strictly 
valid at threshold only, i.e. for $\tau=1$, but there is no 
theoretical argument which justifies its validity at any other momentum 
transfer, where $q^2\neq 4m^2$. 

The $|G_M|^2$ values depend, in principle, on the kinematics where the 
measurement was performed and the angular 
range of integration, however it turns out that these two assumptions for $G_E$ 
lead to comparable values for $|G_M|$.

In the SL region the situation is different. The cross section for the elastic 
scattering 
of electron on protons is sufficiently large to allow the measurements of 
angular 
distribution and/or of polarization observables. The existing data on 
$G_M$ show a dipole behavior  up to the highest 
measured value, $-q^2\simeq$ 31 GeV$^2$ \cite{Ar86} according to
\begin{equation}
G_M(q^2)/\mu_p=G_d,~\mbox{with}~
G_d=\displaystyle\frac{1}
{\left [1-\displaystyle\frac{q^2}{ m_d^2 }\right ]^2},~m_d^2=0.71~\mbox{GeV}^2.
\end{equation}
It should be noticed that the independent determination of both $G_M$ and $G_E$ FFs, from the unpolarized $\ell^-+p$-cross section, has been done up to $-q^2=$ 8.7 GeV$^2$ \cite{And94}, and the further extraction of $G_M$ \cite{Ar86} assumes $G_E=G_M/\mu_p$.
The behavior of $G_E$, deduced from polarization experiment $p(\vec e,e'\vec p)$ differs from $G_M/\mu_p$,
with a deviation from $G_d$ up to 70\% at $-q^2$=5.6 GeV$^2$ \cite{Ga02}. This is the maximum momentum at which new, precise data are available. Extension of this measurement at $-q^2$=9 GeV$^2$ is under way at JLab \cite{00111}.

The main experimental results, which have to be understood, are the following:

\begin{itemize}

\item It has been found that the electric and the magnetic distribution in the proton, at small distances, are not equal, and the the electric charge distribution do not follow a dipole behavior,  as a function of the momentum transfer, as previously assumed.

\item  
The values of $G_M$ in the TL region, 
obtained under the assumption  that $|G_E|=|G_M|$, are larger 
than the 
corresponding SL values. This has been 
considered 
as a proof of the non applicability of the Phr\`agmen-Lindel\"of theorem, ( up to $s$=18 GeV$^2$, at least, or as an evidence that the asymptotic regime is not reached \cite{Bi93}.  Note that in TL region, one uses the notation $s=q^2$, $s=2m^2+2Em$, $E$ is the energy of the antiproton beam (in the LAB system of $\overline{p}+p$-collisions).

\end{itemize}

One can express the angular dependence of the differential cross section for $\overline{p}+p\to \ell^+ +\ell^-$ as a 
function of the angular asymmetry ${\cal R}$ as:
\begin{equation}
\displaystyle\frac{d\sigma}{d(\cos\theta)}=
\sigma_0\left [ 1+{\cal R} \cos^2\theta \right ],~
{\cal R}=\displaystyle\frac{\tau|G_M|^2-|G_E|^2}{\tau|G_M|^2+|G_E|^2}
\label{eq:eq3}
\end{equation}
where $\sigma_0$ is the value of the differential cross section at 
$\theta=\pi/2$. 
These quantities are very sensitive to the different underlying 
assumptions about the $s$-dependence of the FFs \cite{Ia03}, therefore a precise measurement of the ratio ${\cal R}$ would be very interesting. 

The measurement of the differential 
cross section for the process $\overline{p}+p\to \ell^+ +\ell^-$ at a fixed value of 
$s$ 
and for two different angles $\theta$,  allowing  the separation of the two FFs, $|G_M|^2$ and $|G_E|^2$, is equivalent to the well known Rosenbluth separation 
for 
the elastic $ep$-scattering. However in TL, this procedure is simpler, as it 
requires to change only one kinematical variable, $\cos\theta$, whereas, in SL 
it is 
necessary to change simultaneously two kinematical variables: the energy of the 
initial electron and the electron scattering angle, fixing the momentum transfer square, $q^2$. 

The angular dependence of the cross section, Eq. \ref{eq:eq3}, results 
directly from the assumption of one-photon exchange, where the spin of the 
photon 
is equal 1 and the electromagnetic hadron interaction satisfies the 
$C-$invariance. 
Therefore the measurement of the differential 
cross section at three angles (or more) would also allow to test the presence of 
$2\gamma$ exchange \cite{Re99}.

Polarization phenomena will be especially interesting in $\overline{p}+p\to \ell^+ +\ell^-$. For example, the transverse polarization, $P_y$ of proton target (or transverse polarization of antiproton beam) results in nonzero analyzing power \cite{Zi62,Bi93}:
$$
\displaystyle\frac{d\sigma}{d\Omega}(P_y)=
\left ( \displaystyle\frac{d\sigma}{d\Omega} \right )_0 \left [1+{\cal A}P_y
\right ],
$$
$${\cal A}=\displaystyle\frac{\sin 2\theta Im G_E^*G_M}{D\sqrt{\tau}},
~D=|G_M|^2(1+\cos^2\theta)+\displaystyle\frac{1}{\tau}|G_E|^2\sin^2\theta$$
This analyzing power characterizes the T-odd correlation $\vec P\cdot\vec k\times\vec p$, where $\vec k(\vec p)$ is the three momentum of the $\overline{p}$ beam (produced lepton). It is important to note that the $\tau$-dependence of ${\cal A}$ is very sensitive to existing models of the nucleon FFs, which reproduce equally well the data in SL region \cite{Br03}.

The same information can be obtained from the final polarization in $\ell^++\ell^- \to \vec p+\overline{p}$, but in this case one has to deal with the problem of hadron polarimetry, in conditions of very small cross sections.

The main problems in view of a global interpretation of the four nucleon FFs (electric and magnetic, for neutron and proton) in TL and SL momentum transfer region that can be solved by future measurements with a polarized antiproton beam (or with unpolarized antiproton beam on a polarized proton target) are:

\begin{itemize}

\item The separation of the electric and magnetic FFs, through the angular distribution of the produced leptons: the measurement of the asymmetry ${\cal A}$ (from the angular dependence 
of the 
differential cross section for $\overline{p}+p\leftrightarrow \ell^+ +\ell^-$) is 
sensitive to the relative value of $G_M$ and $G_E$.

\item The presence of a large relative phase of magnetic and electric proton FFs 
in 
the TL region, if experimentally proved at relatively large momentum transfer, 
can be considered a strong  
indication that these FFs have a different behavior.

\item The study of the processes $\overline{p}+p\to \pi^0+ \ell^+ +\ell^-$ and $\overline{p}+p\to \pi^++\pi^-+\ell^+ +\ell^-$, will allow to measure proton FFs in the TL region, for $s\le 4m^2$, where the vector meson contribution plays an important role.
\end{itemize}

\vspace*{0.1 true cm}
\noindent\underline{\bf Evaluation of counting rates}
\vspace*{0.1 true cm}

In the hypothesis $|G_M|=|G_E|$, the measured form factors, in the TL region \cite{An03}, can be fitted by the QCD-inspired function: 
$$|G_M|=\displaystyle\frac{56}{s^2}\left( ln\displaystyle\frac{s}{\Lambda^2}\right )^{-2}, \mbox{s~is~expressed~ in~ GeV}^2 $$
with $\Lambda=0.3$ GeV.
After $\cos\theta$ integration, Eq. (\ref{eq:eqcs}) gives:
\begin{equation}
d\sigma(p\overline{p}\to \ell^+\ell^-)=
\displaystyle\frac{4\pi\alpha^2}{3\beta_p s}
|G_M|^2\left (1+\displaystyle\frac{2m^2}{s}\right ),
\label{eq:eqcs1}
\end{equation}
and $\beta_p=\sqrt{1-4m^2/s}$.

So, assuming a luminosity ${\cal L}=2\cdot 10^{32}$ cm$^2$/s \cite{GSI}, we find the number of events per day (1 day=$10^5$ s) as in Table 1.
\begin{center}
\begin{table}[h]
\begin{tabular}{|c|c|c|c|}
\hline
$s$ [GeV $^2$]  & $E $ [GeV]  & $\sigma$ [pb] & Events/day\\
\hline
5  & 2.2   & 830& 16600       \\  
10 & 4.9& 8. & 161      \\  
15 & 7.5& 0.7 & 13      \\
20 & 10.2& 0.12 & 2.4    \\
25 & 12.9& 0.03 & 0.64  \\
30 & 15.5& 0.01 & 0.22\\
\hline
\end{tabular}
\caption[]{Expected counting rates, for $\overline{p}+{p}\to \ell^++\ell^-$.} 
\label{table:kin}
\end{table}
\end{center}
These numbers show that it will be possible, at the future GSI facility, to separe the electric and magnetic FFs in a wide region of $s$ and to extend the measurement of FFs up to the largest available energy, corresponding to $s\simeq 30$ GeV$^2$. 

In conclusion, let us summarize the main points of this note concerning new possibilities opened by the future facility with antiproton beams at GSI:
\begin{itemize}
\item extension of measurements of proton FFs, in the TL region, up to $s=30$ GeV$^2$ (comparable to the maximum value of $-q^2$, achieved in SL region);
\item measurement of the angular dependence of the differential cross section for $\overline{p}+p\to \pi^++\pi^-+\ell^+ +\ell^-$, which will firstly allow the separation of electric and magnetic FFs, in TL region, in a wide region of $s$;
\item using polarized target (or beam) the measurement of the analyzing power will allow to measure for the first time the $s$-dependence of the relative phase of electric and magnetic FFs.
\end{itemize}
This program is especially interesting with respect to the important problem of  the transition to the asymptotic behaviour of nucleon electromagnetic FFs  predicted by QCD, which actually gives rise to many discussions and speculations.

{}

\end{document}